\documentstyle[a4,epsfig,psfig,rotating,subfigure]{article}


\begin{document}

{\large

\title{\bf Performance of the Electromagnetic Calorimeter of the HERMES 
           Experiment}

\author{H. Avakian, N. Bianchi, G.P. Capitani, E. De Sanctis, \\
P. Di Nezza, A. Fantoni$^*$, V. Giourdjian, R. Mozzetti, \\
V. Muccifora, M. Nupieri, A.R. Reolon, P. Rossi\\
{\em  LNF-INFN,
C.P.13, Via E. Fermi 40}\\
{\em I-00044 Frascati (Roma), Italy}\\
J.F.J. van den Brand$^{1,2}$, M. Doets$^2$, T. Henkes$^2$, M. Kolstein$^2$ \\
{\em $^1$Vrije Universiteit Amsterdam, de Boelelaan 1081} \\
{\em 1083 HV Amsterdam, The Nederlands} \\ 
{\em $^2$NIKHEF, Kruislaan 411, 1098 SJ Amsterdam, The Nederlands} \\
A. Airapetian, N. Akopov, M. Amarian, R. Avakian, \\
A. Avetissian, V. Garibian, S. Taroian \\
{\em YPI,
Alikhanian Brothers St. 2, Yerevan AM-375036, Armenia} \\
}

\date{March 9, 1998}
\maketitle 

\begin{abstract}
The performance of the electromagnetic calorimeter of the HERMES experiment
is described.
The calorimeter consists of 840 radiation resistant F101 lead-glass counters.
The response to positrons up to 27.5 GeV, the comparison
between the measured energy and the momentum reconstructed from tracking, 
long-term stability, hadron rejection and neutral meson 
invariant mass reconstruction are shown.
\end{abstract}

{\footnotesize $^*$ 
Corresponding author: e-mail Alessandra.Fantoni@lnf.infn.it}

\newpage

\section{Introduction}
\indent
\par
HERMES ({\bf HER}A {\bf ME}asurement of {\bf S}pin) is an experiment 
which is comprehensively studying the spin structure of the nucleon 
by deep inelastic scattering (DIS) of polarised positrons from polarised 
protons and neutrons \cite{PR}.
Both inclusive and semi-inclusive spin dependent scattering
are simultaneously measured with good particle identification.
\par
By measuring the longitudinal polarization asymmetry of the cross section,
HERMES determines the nucleon spin structure functions in a wide range of $x$ 
and $Q^2$ (0.02 $<x<$ 0.8, 0.2 $<Q^2<$ 20), to more precisely test
fundamental sum rules such as those of Bj\o rken and Ellis-Jaffe.
A central aspect of the physics program is `flavor-tagging' the struck quark
via detection of the leading hadron in semi-inclusive channels, which enables
HERMES to disentangle the spin contributions of different quark flavors and of 
gluons, in an effort to solve the nucleon spin puzzle \cite{SP}. 
\par
The HERMES spectrometer \cite{TDR} is installed in the East Hall of the HERA 
storage ring at DESY. It consists of two identical halves above and below the 
positron ring plane.
This provides two independent measurements of spin observables and thus a 
cross check on systematic uncertainties. 
The spectrometer is configured around a large dipole 
magnet with a bending strength of 1.3 T$\cdot$m and scattering angle 
acceptance 40 -- 220~mrad, a tracking system with chambers before,
in and behind the magnet, and a particle identification detector (PID) system.
The PID system consists of four detectors: a lead-glass calorimeter,
two plastic scintillator hodoscopes, a transition radiation detector, and a 
threshold \v{C}erenkov detector. 
The hodoscope immediately in front of the calorimeter is preceded by two
radiation lengths of lead and acts as a pre-shower detector.
A more detailed description of the spectrometer and of its performances 
is given in Refs. \cite{TDR,MD}.
\indent
\par
The HERMES spectrometer has been in operation for about three years, 
for measurements on polarised and unpolarised targets of $^1$H, $^2$H, $^3$He
and N.  
This paper reports on the performance of the HERMES calorimeter during this 
running period.

\section{Description of the calorimeter}
\subsection{Detector assembly}
\indent
\par
The electromagnetic calorimeter is one of the four detectors of the HERMES PID 
system. Its function is: {\it i)} to provide a first-level 
trigger for scattered positrons, based on  energy deposition in a 
localized spatial region; {\it ii)} to separate positrons from pions with a 
rejection factor of more than 10 at the first-level trigger and an additional 
factor of more than 100 in event reconstruction analysis; {\it iii)} to 
provide a measurement of the energy of DIS positrons; {\it iv)} to measure 
the energy of photons from radiative processes or from $\pi^0$ and $\eta$ 
decays and {\it v)} to give a coarse position measurement of scattered 
electrons and photons.

 \begin{figure} [ht]
  \begin{center}
   \epsfig{file=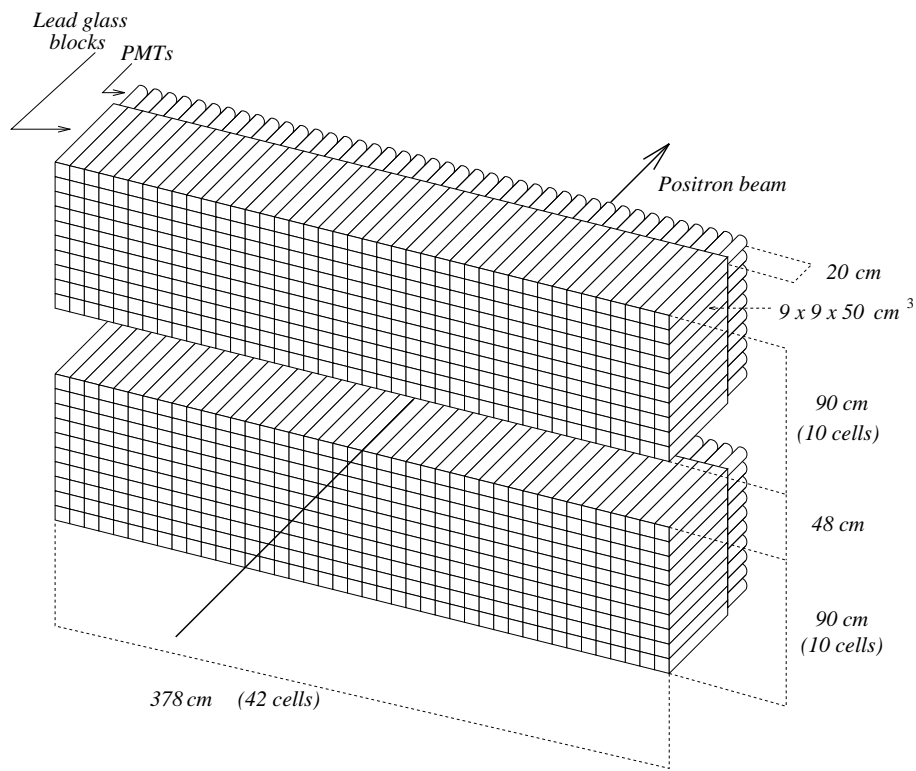,width=12cm}
   \caption{Isometric view of the HERMES calorimeter.}
   \label{fig:calorimeter}
  \end{center}
 \end{figure}

The solution chosen to meet these requirements consists of 840
radiation resistant F101 lead-glass (LG) blocks \cite{RH} arranged in a 
configuration with one wall above and one below the beam, and
with photomultipliers (PMTs) viewing from the rear, as shown in 
Fig. \ref{fig:calorimeter}.
Each wall is composed of 420 identical lead-glass blocks, stacked in a 
42$\times$10 array.
Each block has an area 9$\times$9 cm$^{2}$ and a length of 50 cm 
(about 18 radiation lengths).
This cell size meets the requirement that $\approx$ 90$\%$ of the shower is 
contained in the cell for an axially-incident positron.
The blocks were polished, wrapped with 50 $\mu$m thick aluminized mylar 
foil and covered with a 125 $\mu$m thick tedlar foil to provide light 
isolation.
Each block is coupled to a 7.5 cm photomultiplier Philips XP3461 
with a silicone glue (SILGARD 184) with refraction index 1.41.
A $\mu$-metal magnetic shield of 1.5 mm thickness surrounds the PMT.
The light seal is provided by an aluminium enclosure, which is mounted
on a flange that is glued to the surface of the lead-glass. 
This flange is made of titanium to match the thermal expansion coefficient of 
F101. It carries the light fiber for monitoring of the counter response.
\indent
\par
The characteristics of the F101 blocks were measured at CERN and DESY 
test beams using 3x3 arrays of counters \cite{tesi,NIM}.

\subsection{Equalisation of the counters}
\indent
\par
Before the installation in the HERA East Hall, all lead-glass counters were 
equalised at DESY with a 3 GeV electron beam.
An array of forty-two blocks at a time was placed on a platform
that could be moved in both the horizontal and vertical directions 
to vary the impact point of the beam on the counters. 
The response equalisation procedure consisted in adjusting the PMT 
high voltages so that the mean charge measured by the ADC was
$Q_0$ (pC) = 22.22 $E$ (GeV), where $E$ is the mean energy deposition in the 
cell.
\indent
\par
Figs. 2$a)$ and 2$b)$ show the distributions of the means and variances
(in ADC values) of the spectra of the 840 blocks in response to a 3 GeV 
electron beam incident at the center of the block. 
The mean ADC channels M of all F101 counters were adjusted to be between 
580 and 620. The resulting distribution of the means has a average value
601 and width ($\sigma$) 6.
This means an overall equalisation within 1\%. 
The standard deviations $\Sigma$ of the responses of the 840 lead-glass 
blocks are distributed around a central value 62 with $\sigma$=3: this 
implies a uniformity of the single-block resolutions to within 5\%.

\begin{figure}
 \begin{center}
    \begin{tabular}[t]{c}
      \subfigure[]{\epsfig{figure=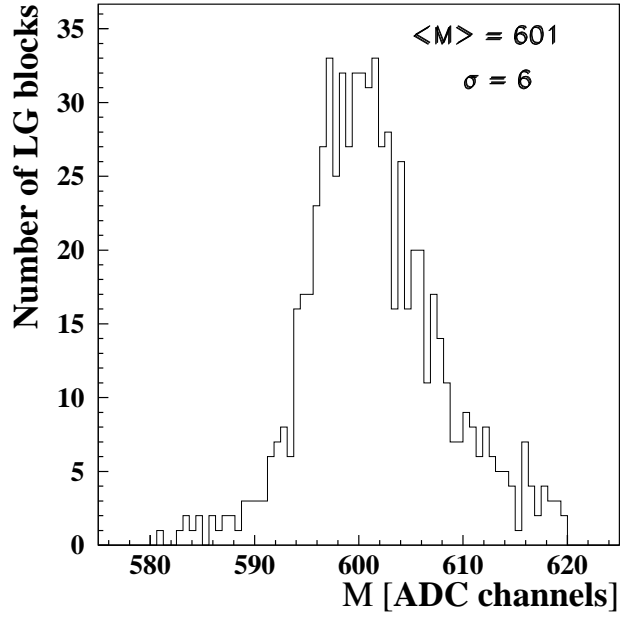,width=9cm}}    \\
      \subfigure[]{\epsfig{figure=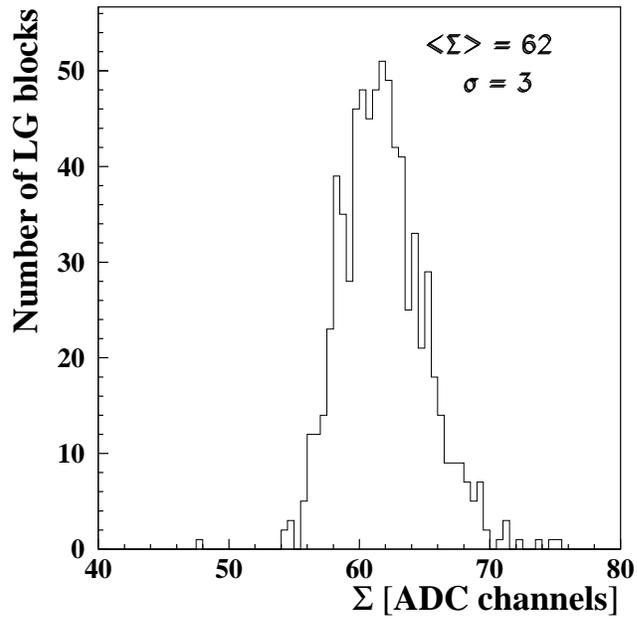,width=9cm}}  
    \end{tabular}
 \label{fig:equalization}
\caption{Equalisation of the 840 lead-glass (LG) blocks in a 3 GeV electron 
         beam: a) Distribution of the mean M ADC values; b) Distribution 
	 of the standard deviation $\Sigma$ of ADC values.}
  \end{center}
\end{figure}

\subsection{Energy calibration}
\indent
\par
The block size was chosen in order to provide containment in a 3x3 matrix of 
more than 99$\%$ of electromagnetic showers up to 30 GeV energy.
Hence shower leakage has negligible influence on the energy resolution. 
On the other hand, the length of the lead-glass module 
does not lead to excessive absorption of \v{C}erenkov light.
\par
Measurements with 1--30 GeV electron beams have been performed at CERN
and DESY with a 3x3 array of counters: all data, apart from that at 1 GeV, 
are reproduced to better than 1$\%$ by a linear fit \cite{NIM}.
\indent
\par
In the off-line analysis of HERMES data, the comparison of the energy $E$ to 
the independently measured momentum $p$ determined by tracking \cite{wwc} 
provides a good identification of scattered positrons over the whole energy 
range which constitutes a powerful tool for calibration.
In fact, after correction for radiative effects in front of the calorimeter, 
the ratio $E/p$ is expected to be close to unity, independently of the positron
energy. 
Fig. 3 shows the calorimeter response for scattered positrons in 
comparison to the reconstructed momentum.
Good linearity is observed over the full energy range.
\indent
\par
During the data taking period (1995-1997) the $E/p$ distribution for scattered
positrons was regularly observed for each individual counter.
Fig. \ref{fig:ep} shows a distribution of the means of such $E/p$ 
spectra, measured over about a one-year running period.
The ratio is distributed around the central value 1.00 with a width ($\sigma$)
0.01, demonstrating a uniformity of response of the counters around $\sim 1\%$.

\begin{figure} 
 \begin{center}
  \epsfig{figure=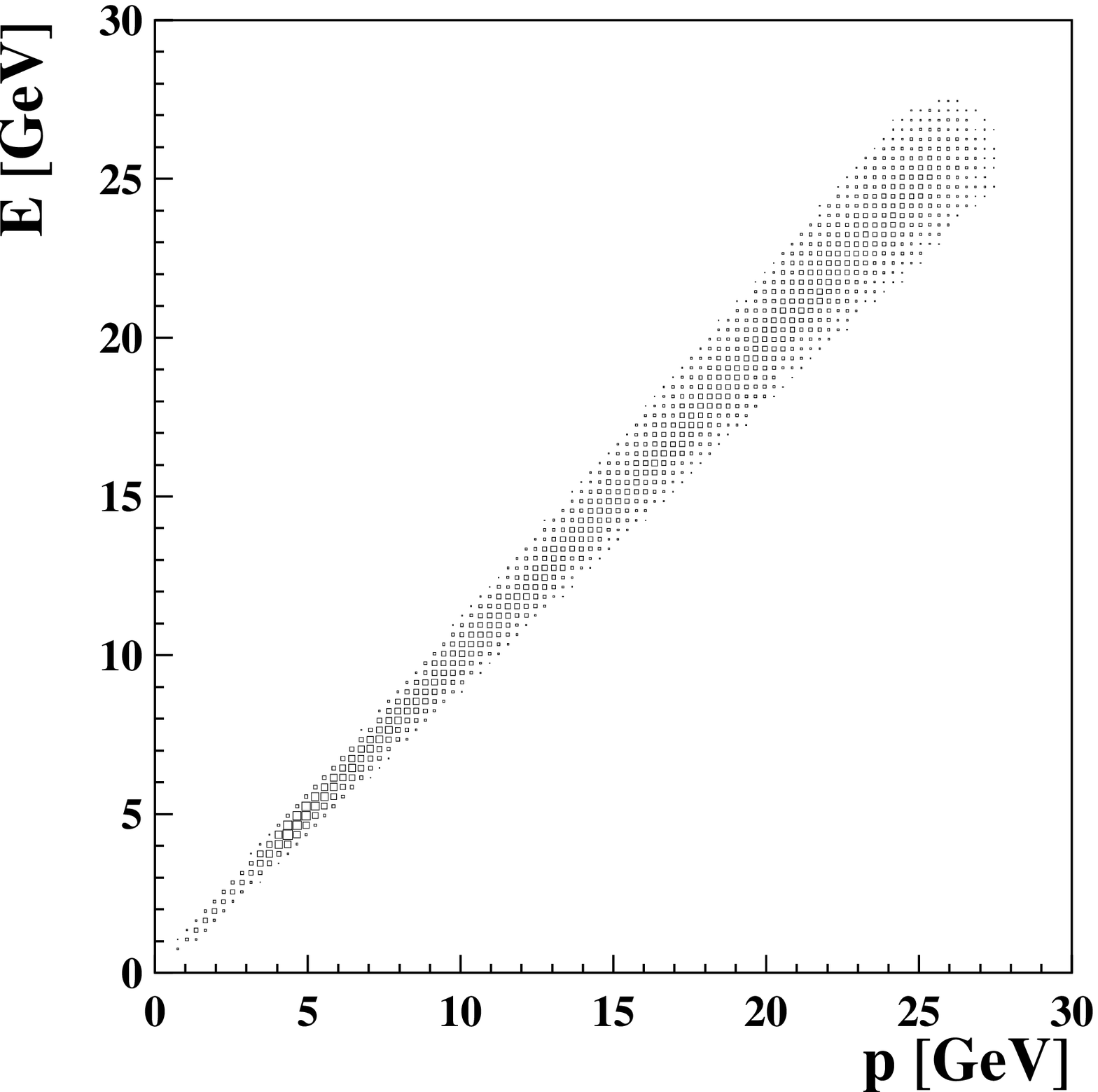,width=9cm}    
  \label{fig:evsp}
  \caption{Positrons energies $E$ measured by the calorimeter versus the 
   positrons momenta $p$ reconstructed in the spectrometer during data taking.}
  \end{center}
  \begin{center}
   \epsfig{file=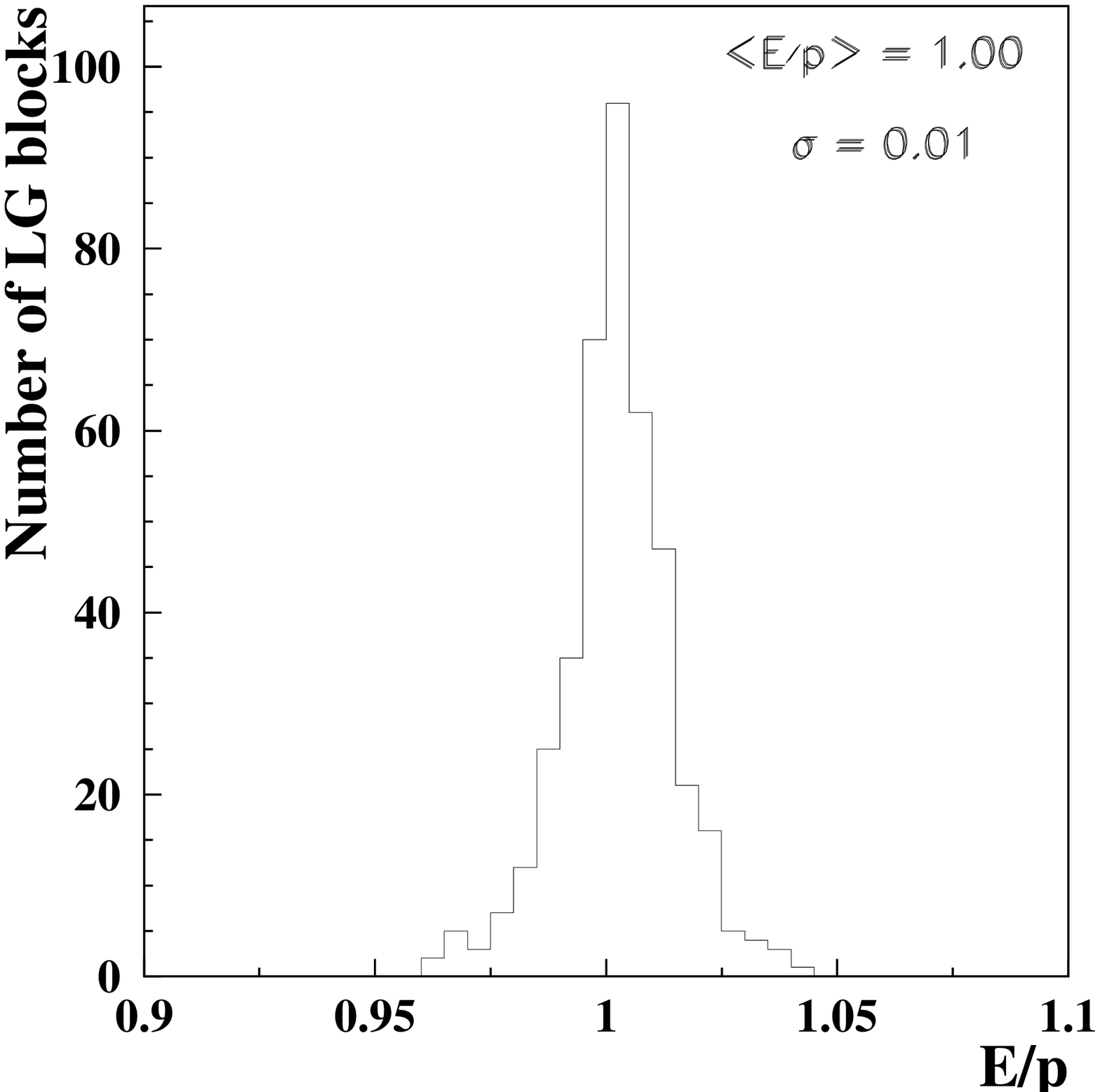,width=9cm}
   \caption{Distribution of the $E/p$ values measured for each counter for all 
    runs collected over one-year data taking period.}
   \label{fig:ep}
  \end{center}
 \end{figure}

\section{Calorimeter performance}
\subsection{Energy resolution}
\indent
\par
Electromagnetic showers typically spread their energy over the eight modules
surrounding the hit counter. Such a group of nine modules is called in the 
following a cluster.
While the energy distribution over the single blocks of the cluster strongly 
depends on the hit position relative to the module boundaries and on the angle 
of incidence, the cluster energy is found to be independent of them to better 
than 1\% \cite{NIM}.
\indent
\par
The energy resolution for scattered positrons obtained during normal operation 
is shown in Fig. \ref{fig:resolution}. The data are well described by the 
following parameterization:
$$
\frac{\sigma (E)}{E} [\%] = \frac{(5.1 \pm 1.1)}{\sqrt{E (\mathrm{GeV})}} + 
(2.0 \pm 0.5) + \frac{(10.0 \pm 2.0)}{E (\mathrm{GeV})}
$$ 
which is slightly degraded compared to the test beam results 
($ \sigma (E) /E [\%] = (5.1 \pm 1.1) / \sqrt{E (\mathrm{GeV})} + 
(1.5 \pm 0.5) $) \cite{NIM}.
This because of pre-showering of the positrons in the material 
before the calorimeter, which improves the discrimination between positrons 
and hadrons, but produces the $E^{-1}$ term, and of imperfections in the gain 
matching among modules, which slightly enhances the constant term.
Note that these values are similar to those obtained for 
other large lead-glass calorimeters [9-15]
in spite of the use here of a less transparent 
material.

\begin{figure}[ht] 
 \begin{center}
   \epsfig{file=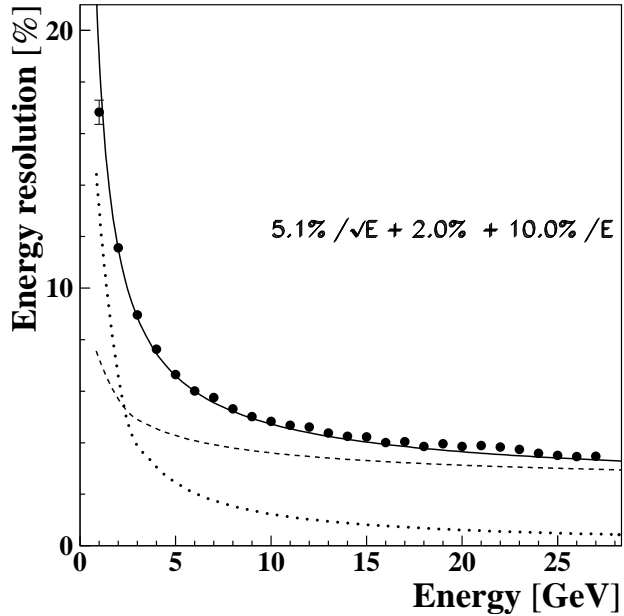,width=9cm}
   \caption{Energy resolution of the calorimeter: the circles correspond to 
	    the data for $E/p$ after subtraction of the resolution 
            contribution for $p$ as predicted by Monte Carlo; the solid curve 
            is the sum of the contributions from
            the lead-glass (dashed curve) and from the preshower (dotted 
            curve) provided at test beam measurement [7].}
   \label{fig:resolution}
 \end{center}
\end{figure}

\subsection{Position resolution}
\indent
\par
The segmentation of the calorimeter allows to obtain the hit position 
from the energy distribution inside a cluster with an accuracy better than
the cell size. The hit position is calculated by using the following 
energy-weighted average position of the nine blocks of a cluster:
$$ x = \frac{\sum_{i=1}^{9} x_i \sqrt{E_i}}{\sum_{i=1}^{9} \sqrt{E_i}} $$
and $$ y = \frac{\sum_{i=1}^{9} y_i \sqrt{E_i}}{\sum_{i=1}^{9} \sqrt{E_i}}, $$
where $x_i$ and $y_i$ are the central horizontal and vertical coordinates of 
the $i$-module and $E_i$ is the corresponding measured amplitude. 
Fig. \ref{fig:position} shows the distribution of the differences 
$\Delta x = x_{\mathrm{calo}} - x_{\mathrm{track}}$ between the estimated hit 
positions of scattered positrons in the calorimeter $x_{\mathrm{calo}}$, 
and the extrapolations of the charged particle tracks, $x_{\mathrm{track}}$ 
\cite{calor96}. 
It is seen that the resolution of the reconstructed hit position amounts to:
$$\sigma_x \approx 7 \mathrm{mm} , $$
and is significantly better than the cell size.
These resolutions, which are the same for the $x$ and $y$ directions, were 
found to be almost independent of the energy $E$ of the incident positron.

 \begin{figure}[ht]
  \begin{center}
   \epsfig{file=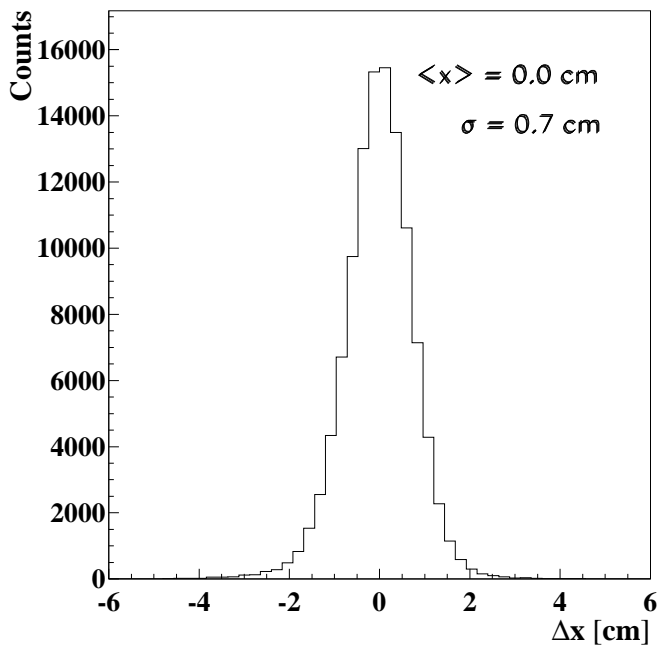,width=9cm}
   \caption{Distribution of the differences $\Delta x = x_{\mathrm{calo}} - 
	    x_{\mathrm{track}}$ between the hit positions measured by the 
	    calorimeter and those determined by the spectrometer.}
 \label{fig:position}
 \end{center}
 \end{figure}

\subsection{Trigger}
\indent
\par
The energy of the electromagnetic shower measured as the sum of two adjacent
calorimeter columns is used to provide the first-level trigger for positrons 
in a deep inelastic process.
During the 1995 data taking period the trigger consisted of a coincidence of 
both hodoscopes and the calorimeter. 
The trigger threshold was set to a deposited energy of 3.5 GeV in 1995 and 
of 1.5 GeV in late 1996 and 1997. 
This already provided a suppression of hadronic background of about
one order of magnitude.  A forward trigger scintillator system was introduced
in 1996 in front of the HERMES spectrometer magnet. It reduced the trigger rate
from background generated by the HERA proton beam by distinguishing 
forward and backward going particles by using the time of flight between 
forward and rear scintillators.

\section{Long-term stability}
\subsection{Gain Monitoring System}
\indent
\par
A gain monitoring system (GMS) is used to monitor the possible gain 
variations of the photomultipliers during normal running. 
The system is based on a dye laser light source at 500 nm, which sends light
pulses of varied intensities through glass fibers to every PMT of the
calorimeter, and additionally to a reference counter photodiode. The different
intensities are achieved by a rotating wheel with several attenuation plates.
The light is split in several stages and fed into glass fibers \cite{TDR}. 
The ratios of multiplier signals to that of the reference photodiode can 
be used to monitor relative gain changes in the multipliers.
\indent
\par
The long-term stability of the calorimeter has been evaluated by observing 
changes of the pedestal and gain value. These values have been found to 
be stable within the accuracy of the measurement during the entire time of 
operation. 
Fig. \ref{fig:piedistalli} shows the values of pedestals 
observed over a several months running period for two typical modules.
From the known conversion gain of 5 MeV/ch, it can be seen that
the data are consistent to about 10 MeV.
\indent
\par
Fig. \ref{fig:gain96} shows the relative gain variations of two typical 
counters as a function of accumulated events for several months running period.
The values are the ratios between the actual and reference gains. 
Straight line fits to the data result in slopes of 1.2$\times$10$^{-3}$ 
year$^{-1}$ and 1.1$\times$10$^{-2}$ year$^{-1}$, confirming the above stated 
long-term stability of the response to within 1\% per year.

\begin{figure} 
 \begin{center}
  \epsfig{file=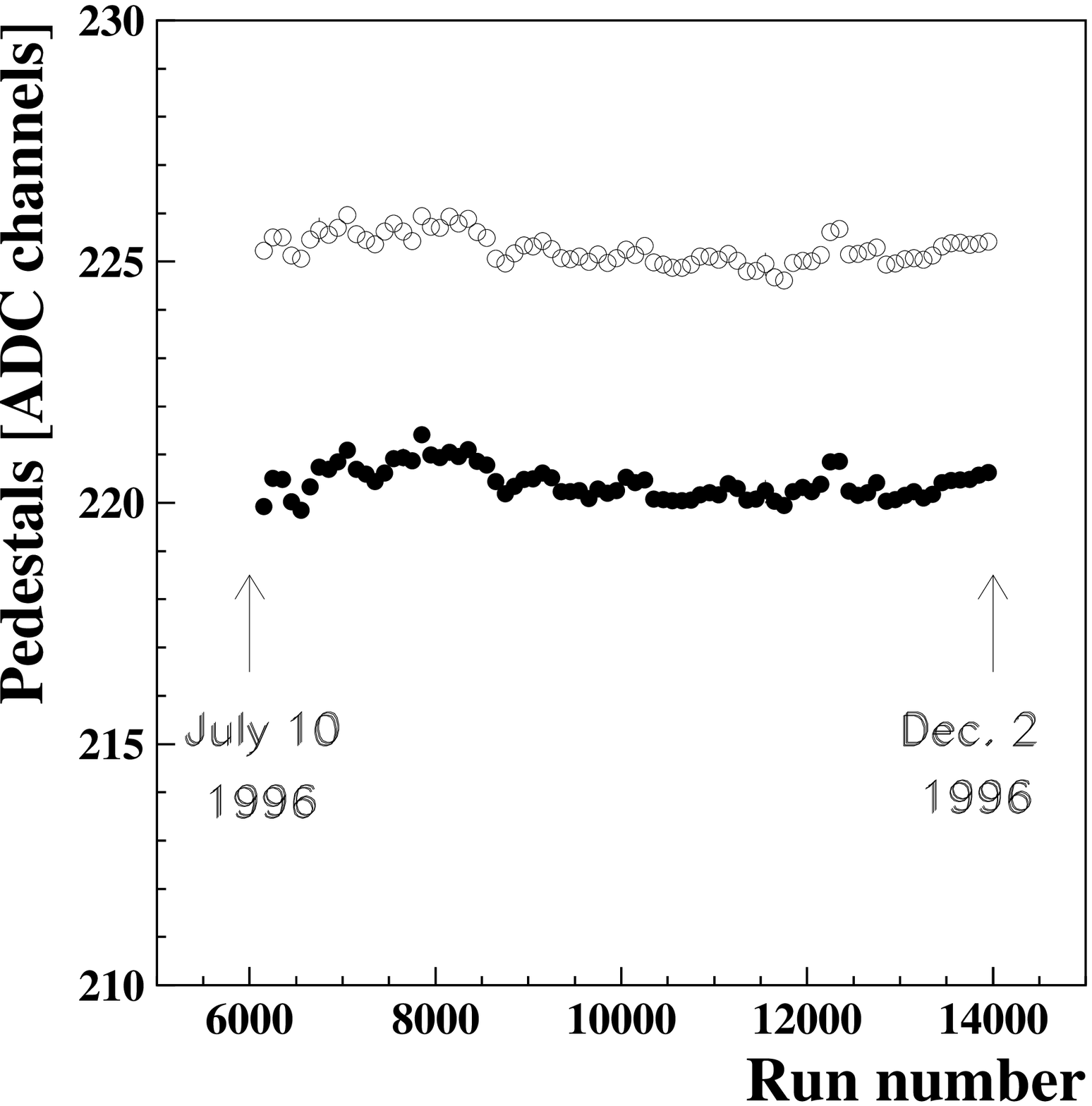,width=9cm}
  \caption{Pedestal behavior for two typical counters during the 1996 
	   running period.} 
  \label{fig:piedistalli}
 \end{center}
 \begin{center}
   \epsfig{file=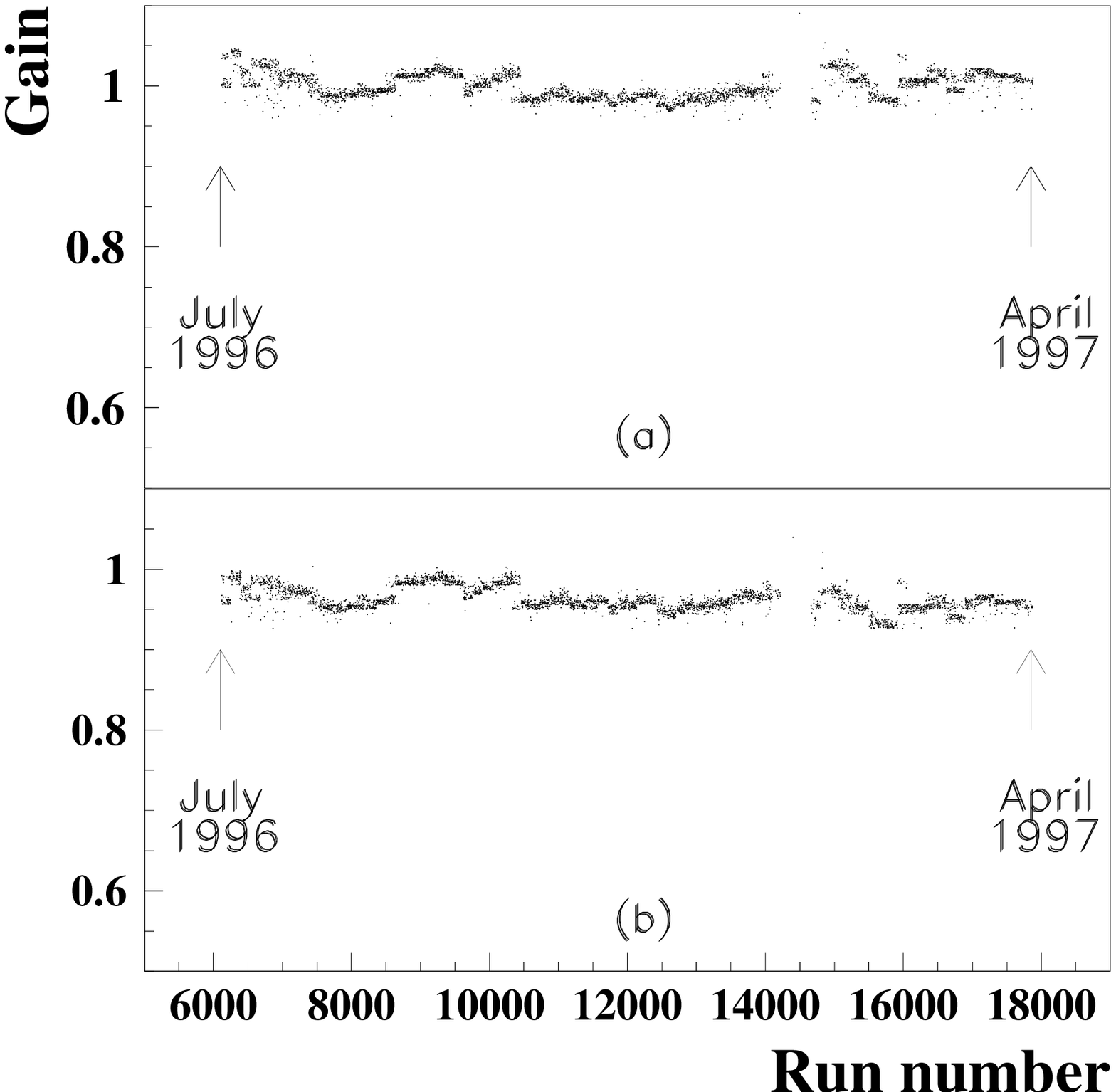,width=9cm}
   \caption{Relative gain variations of two typical PMTs of the lead-glass 
	    counters measured during the 1996 and part of 1997 data taking 
	    periods (a typical length of a run is about 10 min.); 
	    a) counter from top calorimeter wall, b) counter from 
 	    bottom calorimeter wall.}
   \label{fig:gain96}
 \end{center}
\end{figure}

The long-term stability of the response can also be monitored 
by observing the mean value of the $E/p$ distribution, measured for each run.
Fig. \ref{fig:epmeanrun} shows the distribution of the averages over all
blocks of their $E/p$ centroids, accumulated over a one year running period. 
It is seen that the response was stable within 0.5$\%$ (corresponding to the 
$\sigma$ of the distribution).

\begin{figure}[ht] 
 \begin{center}
   \epsfig{file=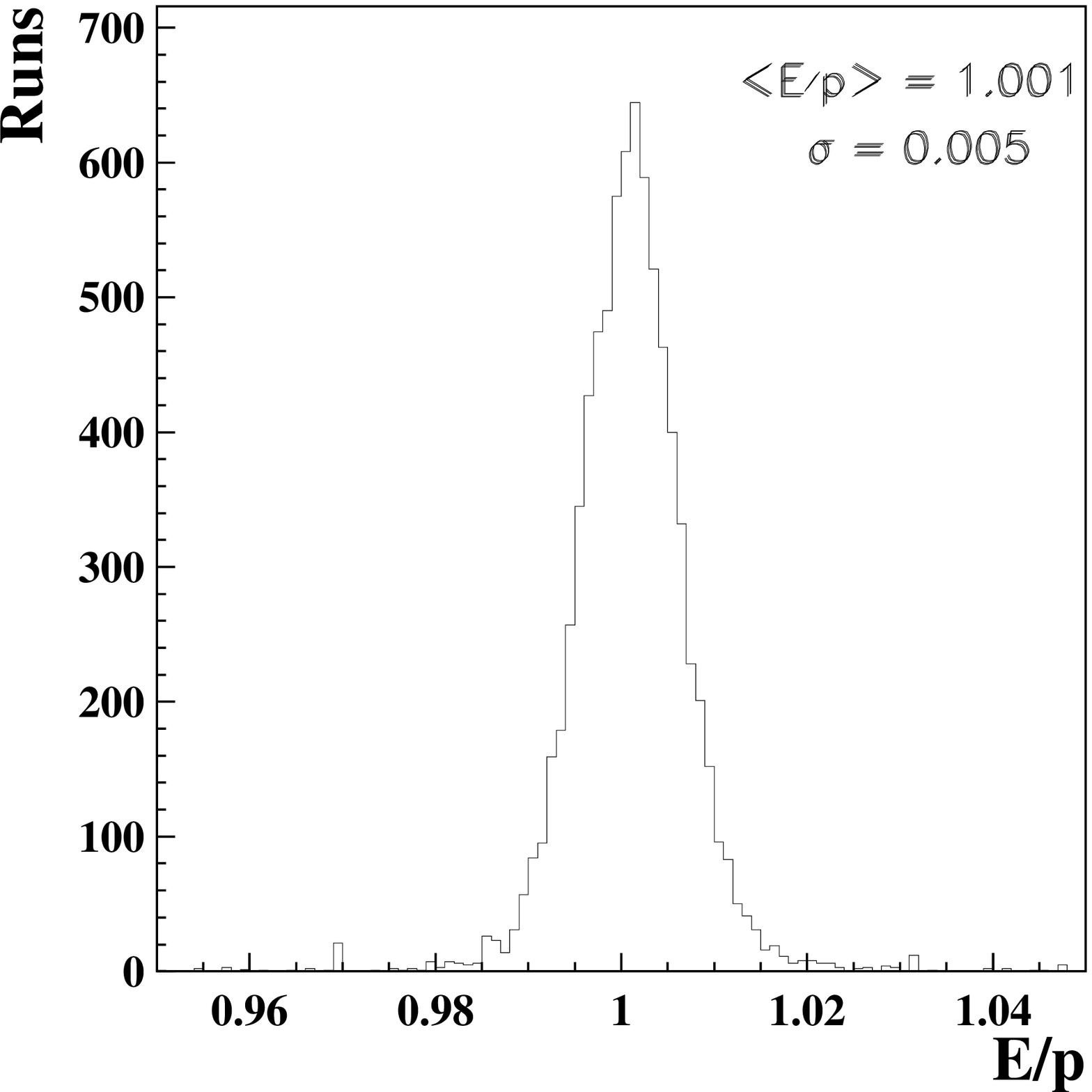,width=9cm}
   \caption{$E/p$ averaged distribution for all lead-glass blocks over a
            one-year of data taking period.}
   \label{fig:epmeanrun}
 \end{center}
\end{figure}

\subsection{Radiation damage}
\indent
\par
Degradation of the optical properties of the lead-glass by radiation is a 
danger in the HERA environment.
The choice of F101 material was motivated by its radiation hardness.
In fact, previous measurements on a 45 cm long block with $\gamma$ rays 
\cite{RH} and high-energy hadrons \cite{IN} have shown that an accumulated 
dose of more than 10$^2$ Gy produces a degradation of F101 transmittance less 
than $1/e$ over the lead-glass length.
After irradiation by 10$^4$ Gy the F101 turned visibly rust-brown with a
tint of red and did not recover.
Thus F101 is expected to be 10--50 times less sensitive to radiation damage 
than other types of lead-glass, like SF2 \cite{HO}, depending on wavelength.
This is due to the addition of Cerium, which has the disadvantage that it 
worsens the optical transmission characteristics.
\par
To prevent radiation damage of the lead-glass, both calorimeter walls are 
vertically displaced away from the beam pipe by 50 cm during beam injection.
Therefore, to monitor the potential radiation damage, particular attention was 
devoted to those blocks positioned at lower scattering angles, which should 
suffer a stronger gain reduction due to their proximity to the beam. 
Fig. \ref{fig:gain96.top.bottom} shows the distributions of the 
relative gains for a few of these blocks measured during one year of operation.
The central values are at 1.003 and 0.9999 for the top and bottom 
walls, respectively, with a $\sigma$ $\approx$ 1$\%$.
This result is also confirmed by the long-term stability of the response to 
the GMS pulses shown in Fig. \ref{fig:gain96}. 
\indent
\par
In conclusion, over three years of operations, there has been no observed 
degradation of performance that could suggest ageing effects.

\begin{figure}[ht] 
 \begin{center}
   \epsfig{file=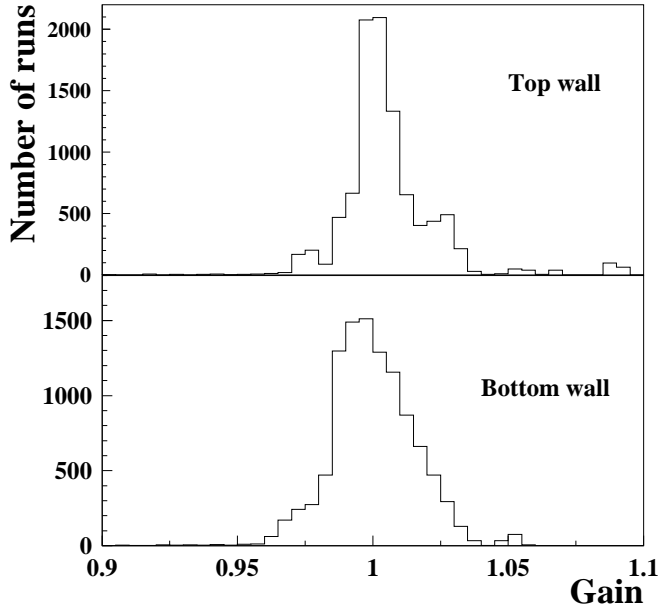,width=9cm}
   \caption{Distribution of the relative gain from the GMS for calorimeter 
	    blocks located in the vicinity of the beam.}
   \label{fig:gain96.top.bottom}
 \end{center}
\end{figure}

Radiation damage to the lead-glass is also monitored by using dedicated TF1 
blocks placed behind the calorimeter. 
This material is about 20 times more sensitive to radiation damage than F101 
\cite{RH}. 
Therefore, gain reduction would be seen sooner in these monitor 
detectors if there had been a large radiation dose incident on the back of the 
calorimeter caused by showers produced by beam loss in the HERA proton 
storage ring.
Within the reproducibility of the measurements (1\%), no variation has yet
been observed in their response, indicating that the effect of radiation 
damage is negligible.

\section{Hadron rejection}
\indent
\par
The HERMES PID system has been designed to provide at least an order of 
magnitude in hadron suppression at the trigger level to keep data acquisition
rates reasonable, and to provide a hadron rejection factor (HRF) of 10$^4$ to 
keep the contamination of the positron sample by hadrons below 1\% over the 
entire kinematic range. 
The HRF is defined as the ratio of the total number of incident hadrons to the 
number of hadrons that are misidentified as electrons.
\indent
\par
The calorimeter and the hodoscopes are used to select DIS events.
This selection is more critical near the energy threshold where 
the ratio between the fluxes of pions and positrons is high. It is 
accomplished with a passive radiator composed of 2 radiation lengths of lead 
sandwiched between two 1.3 mm stainless steel sheets and installed immediately
before the second hodoscope. 
This passive radiator acts like a preshower and initiates electromagnetic 
showers that deposit significantly more energy in the scintillator than 
minimum ionizing particles. 
\par
Measurements with test beams \cite{NIM} have shown that such a 
configuration yields a hadron rejection factor of $\approx$ 5$\times$10$^{3}$
in an event reconstruction analysis combining a lead-glass cut retaining 
95\% electron efficiency with a preshower cut keeping 98\%.
Specifically, the pion rejection provided by a single lead-glass block is 
about 100 and this is improved by the preshower by a factor of about 40.
\indent
\par
During data acquisition the hadron contamination at the trigger level was
suppressed by the calorimeter threshold by a factor 10--100, depending 
on positron energy and threshold setting. 
\indent
\par
In Fig. \ref{fig:rej_eff} are shown the additional HRF and the efficiency for 
the combined calorimeter+preshower system obtainable in off-line analysis:
the HRF (efficiency) values increase from $\sim$50 (0.94) at 4.5 GeV up to 
$\sim$160 (0.98) at 13.5 GeV.
It's worth noticing that in the event reconstruction the responses of the four 
PID detectors (electromagnetic calorimeter, pre-shower, \v{C}erenkov counter,
and transition radiator detector (TRD) \cite{TRD}) are combined to further
improve the hadron rejection to the required value.
More detailed studies on the particle identification system can be found in 
Ref. \cite{PID}.

\begin{figure}[ht] 
 \begin{center}
  \epsfig{file=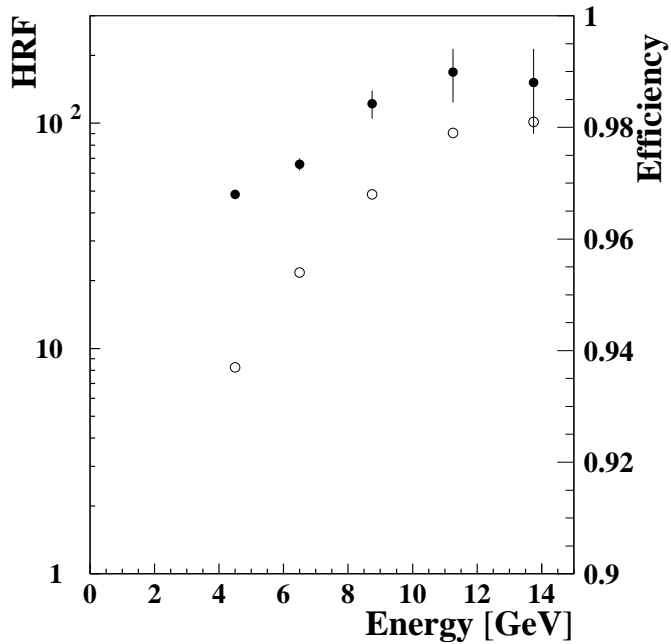,width=9cm}
   \caption{Off-line hadron rejection factors for the system 
	    calorimeter+preshower (full circles and left scale) and 
	    corresponding efficiencies (empty circles and right scale).}
 \label{fig:rej_eff}
 \end{center}
\end{figure}

\section{Invariant mass reconstruction}
\indent
\par
HERMES provides detailed information on the hadronic final states 
in semi-inclusive deep inelastic scattering measurements. 
This yields information on various flavor contributions to the nucleon 
spin.
\indent
\par
The calorimeter plays an essential role in the identification of $\pi^0$ 
and $\eta$, because they mainly decay into two photons (branching ratios:
(98.80$\pm$0.03) \% and (39.2$\pm$0.3) \% respectively \cite{PDG}),
which are identified as pairs of energetic clusters in the calorimeter with no 
corresponding charged tracks in the spectrometer.
From the energy measurement of the two photons and the opening angle between 
them, it is possible to reconstruct the invariant mass of the corresponding 
meson.
Fig. 12$a)$ shows an invariant mass distribution for events with two neutral
clusters in the calorimeter in coincidence with a scattered positron. 
Both the $\pi^0$ and $\eta$ peaks are clearly visible.
Fig. 12$b)$ and Fig. 12$c)$ show the $\pi^0$ and $\eta$ invariant masses
distributions obtained after applying kinematical cuts and background 
subtraction.
The centroids of the peaks are $M_{\pi^0}$=0.135 GeV with $\sigma$=0.011 
GeV, and $M_{\eta}$=0.549 GeV with $\sigma$=0.030 GeV, which are in good 
agreement with the Particle Data Group values \cite{PDG}.
\indent
\par
The resolution of the estimate $M$ for the meson invariant mass can be 
expressed as follows:
$$
\frac{\sigma}{M} = \biggr[ \biggr(\frac{\sigma_{E_1}}{2E_1}\biggr)^2 + 
\biggr(\frac{\sigma_{E_2}}{2E_2}\biggr)^2 + 
\biggr(\frac{\sigma_{\varphi}}{2 {\mathrm{tan}} (\varphi/2)}\biggr)^2 
\biggr]^{1/2}
$$
where $\varphi$ is the opening angle between the two photons.
Using this equation we can examine whether the energy and 
position resolutions derived from calibration data still apply in the
experimental environment. 
Fig. \ref{fig:resolpi0} displays the $\pi^0$ invariant mass resolution 
obtained from DIS events, and a Monte Carlo calculation based on the
measured energy and position resolutions given in section 3. 
There is good agreement between the measured and calculated 
values. 
At low energies the energy resolution dominates the invariant mass resolution, 
while at high energies the angular resolution is more important.
The decrease of slope for energies $\geq$ 9 GeV is due to a cut on minimum 
inter-cluster distance related to the cell size.

\begin{figure}
 \begin{center}
   \begin{tabular}[t]{c}
    \subfigure[]{\epsfig{figure=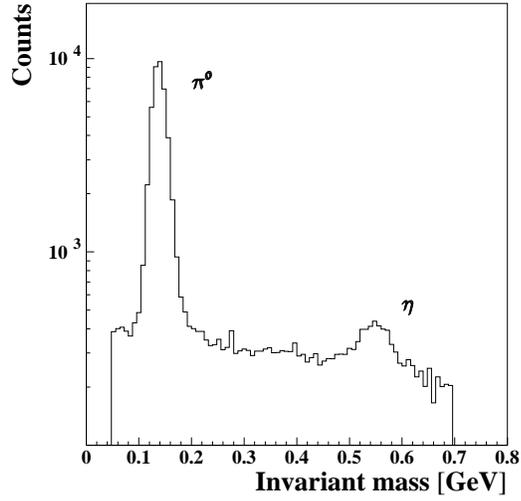,width=7cm}}    \\
    \subfigure[]{\epsfig{figure=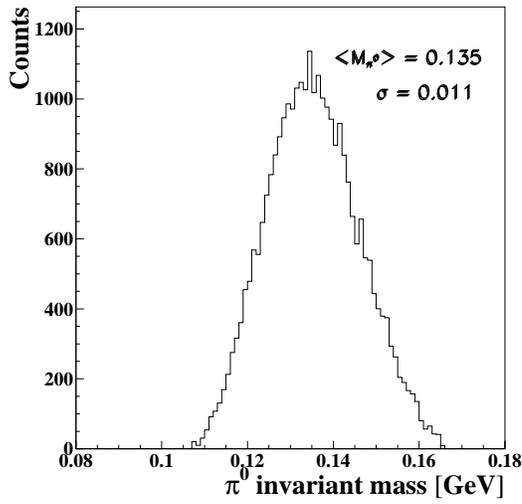,width=7cm}}       
    \subfigure[]{\epsfig{figure=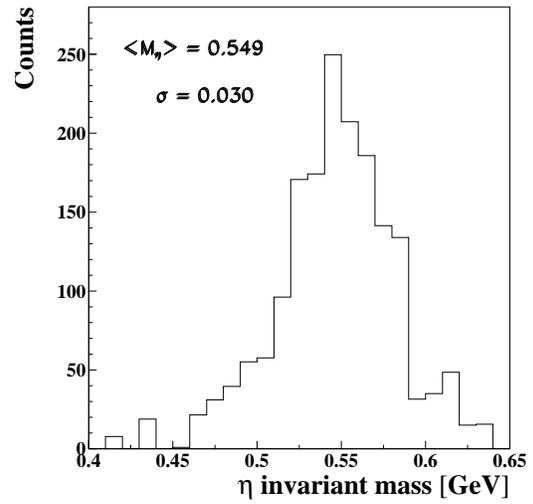,width=7cm}}  
   \end{tabular}
  \label{fig:invmass}
  \caption{Two-photon invariant mass distribution in the calorimeter:
	 (a) peaks of $\pi^0$ and $\eta$; (b) and (c) peaks of $\pi^0$ and of 
         $\eta$ respectively, after applying kinematical cuts and background 
         subtraction.}
 \end{center}
\end{figure}

 \begin{figure}[ht]
  \begin{center}
   \epsfig{file=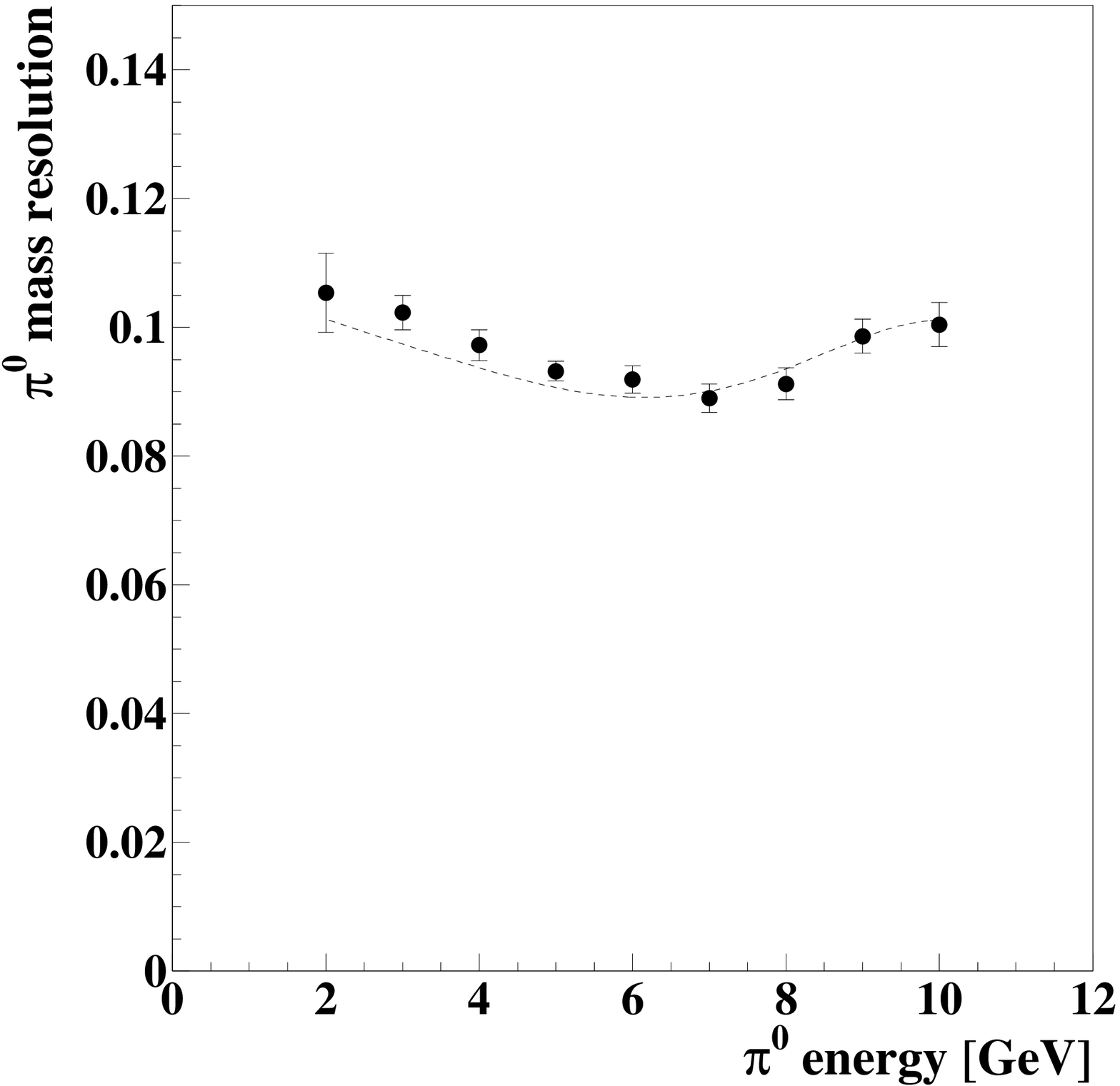,width=9cm}
   \caption{Resolution of the $\pi^0$ invariant mass as calculated from the
            measured energy and position resolutions of two-photon events in 
	    the calorimeter (dashed curve), compared to the values obtained 
	    from semi-inclusive DIS events (closed circles).}
   \label{fig:resolpi0}
  \end{center}
 \end{figure}

\section{Conclusions}
\indent
\par
The electromagnetic calorimeter is an important component of the HERMES 
spectrometer. It provides the DIS trigger of the experiment in conjunction 
with scintillator hodoscopes and plays a major role in the particle 
identification.
In addition, it is essential for the identification of neutral particles in 
semi-inclusive measurements.
The performance and the stability of the calorimeter response were 
continuously measured during the past three years of data taking and the
data are in good agreement with the design values and expectations.
They can be summarized as follows:
\begin{itemize}
\item uniformity of the response of all counters within 1$\%$;
\item linearity of the response to positrons within 1$\%$ over the energy
range 1--30 GeV;
\item resolution
$$ \frac{\sigma (E)}{E} [\%] = \frac{(5.1 \pm 1.1)}{\sqrt{E (\mathrm{GeV})}} + 
(1.5 \pm 0.5) $$
for a 3x3 array of counters and
$$ \frac{\sigma (E)}{E} [\%] = \frac{(5.1 \pm 1.1)}{\sqrt{E (\mathrm{GeV})}} + 
(2.0 \pm 0.5) + \frac{(10.0 \pm 2.0)}{E (\mathrm{GeV})}$$ for the whole 
calorimeter operating in the spectrometer, including the effect of 
pre-showering of the positrons in the material before the calorimeter;
\item position reconstruction with resolution about 0.7 cm;
\item stability in time of the response within 1$\%$;
\item no observed degradation of performance due to radiation damage, within 
the accuracy of the measurements;
\item a hadron rejection factor exceeding 10 at the trigger level, and
a further off-line rejection factor of about 100;
\item reconstruction of $\pi^0$ and $\eta$ masses in agreement with the PDG
values.
\end{itemize}

\section*{Acknowledgements}
\indent
\par
We gratefully acknowledge the Frascati HERMES technical staff M. Albicocco, 
A. Orlandi, W. Pesci, G. Serafini and A. Viticchi\'{e} for the continuous and 
effective assistance. We are also grateful to C.~A.~Miller for many
stimulating discussions.
This work was in part supported by the INTAS (contract number 93-1827) and the 
TMR research network (contract number FMRX-CT96-0008) contributions from 
the European Community.

}
\end{document}